\documentclass[12pt,preprint]{aastex}

\begin{document}

\title{Spatial Locality of Galaxy Correlation Function in Phase Space:
  Samples from the 2MASS Extended Source Catalog}

\author{Yicheng Guo\altaffilmark{1}, Yao-Quan Chu\altaffilmark{1} and
                      Li-Zhi Fang\altaffilmark{2}  }

\altaffiltext{1}{Center for Astrophysics, University of Science
and Technology of China, Hefei, Anhui 230026,P.R.China}
\altaffiltext{2}{Department of Physics, University of Arizona,
Tucson, AZ 85721}

\begin{abstract}

We analyze the statistical properties and dynamical implications of
galaxy distributions in phase space for samples selected from the
2MASS Extended Source Catalog. The galaxy distribution is decomposed
into modes $\delta({\bf k, x})$ which describe the number density
perturbations of galaxies in phase space cell given by scale band
$\bf k$ to ${\bf k}+\Delta {\bf k}$ and spatial range $\bf x$ to
${\bf x}+\Delta {\bf x}$. In the nonlinear regime,
$\delta({\bf k, x})$ is highly non-Gaussian. We find, however,
that the correlations between $\delta({\bf k, x})$ and
$\delta({\bf k', x'})$ are always very weak if the spatial ranges
(${\bf x}$, ${\bf x}+\Delta {\bf x}$) and (${\bf x'}$, ${\bf
x'}+\Delta {\bf x'}$) don't overlap. This feature is due to the
fact that the spatial locality of the initial perturbations
is memorized during hierarchical clustering. The highly spatial
locality of the 2MASS galaxy correlations is a strong evidence for
the initial perturbations of the cosmic mass field being spatially
localized, and therefore, consistent with a Gaussian initial
perturbations on scales as small as about 0.1 h$^{-1}$ Mpc. Moreover,
the 2MASS galaxy spatial locality indicates that the relationship
between density perturbations of galaxies and the underlying dark
matter should be localized in phase space. That is, for a structure
consisting of perturbations on scales from $k$ to $ k+\Delta { k}$,
the nonlocal range in the relation between galaxies and dark matter
should {\it not} be larger than $|{\Delta {\bf x}}|=2\pi/|\Delta {\bf k}|$.
The stochasticity and nonlocality of the bias relation between
galaxies and dark matter fields should be no more than the allowed
range given by the uncertainty relation
$|{\Delta {\bf x}|| \Delta{\bf k}}|=2\pi$.

\end{abstract}


\section{Introduction}

The large scale structure of the universe developed from initial
mass density and velocity fluctuations through gravitational
instability. Much of the information of the initial perturbations is
``forgotten" during the gravitational nonlinear evolution. Yet,
some features of the initial perturbations are imprinted in the
cosmic mass and velocity fields at present. Measuring these memorized
features is crucial in studying the initial states of the universe.
A well known example is the self-similarity of gravitational clustering,
which imprints the initial power spectrum index and is detectable
with the scaling behavior of correlation functions of the present mass
field (e.g. Peebles 1980). In this paper, we study the spatial
locality of the perturbed mass field in phase space,
which is also a feature to be memorized in mass field today.

The physics of spatial locality of correlation function in phase space
can be illustrated with a Gaussian mass field $\rho({\bf x})$ described by
\begin{equation}
\langle\hat{\delta}({\bf k}) \hat{\delta}({\bf -k'})\rangle =
P(k)\delta^K_{\bf k,k'},
\end{equation}
where $\hat{\delta}({\bf k})$ is the Fourier counterpart of the
density contrast
$\delta({\bf x})=[\rho({\bf x})-\overline\rho]/\overline{\rho}$,
$P(k)$ the power spectrum of the mass field, and $\delta^K$ the
Kronecker delta function. Eq.(1) says that the Fourier modes with
different wavevector ${\bf k}$ are uncorrelated, or the correlation is
localized in $k$-space. This is because the phase of the Fourier modes
$\delta({\bf k})$ is random. On the
other hand, the correlation function of density perturbations in physical
($x$) space generally is non-local. The two-point correlation function
$\langle \delta({\bf x})\delta({\bf x'})\rangle$ has non-zero correlation
length when the Fourier power spectrum $P(k)$ of eq.(1) is $k$-dependent.

In a phase-space description, the mass field is decomposed into modes
$\delta({\bf k, x})$, the perturbations in the wavevector(scale)
from ${\bf k}$ to ${\bf k} + \Delta {\bf k}$ and physical
range ${\bf x}$ to ${\bf x}+\Delta {\bf x}$. The volume of the phase
space cell referring to a mode is given by the uncertainty relation
$|\Delta {\bf x}||\Delta {\bf k}| =2\pi$. The correlation function of a
Gaussian field generally is localized regardless whether the Fourier power
spectrum is colored. That is,
\begin{equation}
\langle \delta({\bf k, x})\delta({\bf k', x'})\rangle \propto
\delta^K_{\bf x,x'}\delta^K_{\bf k,k'}.
\end{equation}
The reason for eq.(2) is straightforward. First, the perturbations
$\delta({\bf k, x})$ and  $\delta({\bf k', x'})$ are,
respectively, given by linear superposition of the Fourier modes
in different wavebands (${\bf k}$, ${\bf k}+ \Delta {\bf k}$) and
(${\bf k'}$, ${\bf k'}+ \Delta {\bf k'}$). For a Gaussian field,
the Fourier modes in different wave band are uncorrelated in
general [eq.(1)]. This gives rise to the factor of $\delta^K_{\bf
k, k'}$ of eq.(2). Second, the phases of the Fourier modes of
Gaussian field are random. For a superposition of the Fourier
modes in the band (${\bf k}$, ${\bf k} +\Delta {\bf k}$) with
random phase, the coherent length of the phase of $\delta({\bf
k,x})$ is not larger than $|\Delta {\bf x}| \simeq 2\pi/|\Delta
{\bf k}|$. On the other hand, the spatial distance between two
cells in the ($k-x$) space is at least $|\Delta {\bf x}| \simeq
2\pi/|\Delta {\bf k}|$. This yields the factor $\delta^K_{\bf
x,x'}$ of eq.(2). Therefore, the phase space correlation function
eq.(2) for Gaussian field generally is local.

Non-linear evolution via gravitational clustering will lead to a
non-Gaussian field that will deviate from the locality of eq.(2),
even when the initial field
is Gaussian. However, turbulence studies have found that if  1.) the
initial perturbations are spatially localized, and 2.) the random fields
evolve via a self-similar hierarchical cascade process, the phase space
correlation function of the evolved field will still be spatially localized
(Greiner, Lipa, \& Carruthers, 1995, Greiner et al. 1996). The perturbations
of modes $\delta({\bf k, x})$ and $\delta({\bf k', x'})$ with
${\bf x}\neq {\bf x'}$ stay statistically uncorrelated or only
very weakly correlated during the self-similar hierarchical cascade
evolution. That is, the factor $\delta^K_{\bf x,x'}$ of eq.(2) is memorized
during the dynamical evolution.

Phenomenological models that mimic the hierarchical clustering of
the cosmic mass field have similar mathematical structures as the
hierarchical cascade models of turbulence. For instance, the
fractal hierarchy clustering model of Soneira \& Peebles (1977) is
the same as the $\beta$ model of turbulence (e.g. Frisch, 1995).
The block model of Cole \& Kaiser (1988) is a special case of the
multifractal cascade model (Meneveau \& Sreenivasan 1987, Pando et
al 1998). Hence, these models should also memorize the spatial
locality. Recently, the spatial locality has been studied with
more realistic dynamical models of gravitational clustering.
First, if the weakly nonlinear mass field is given by the
Zeldovich approximation, the field is found to be spatially
localized if the initial perturbations are Gaussian (Pando, Feng
\& Fang 2001). More recently, this result has been extended to
fully nonlinear regime (Feng \& Fang 2004). Using the halo model
of the large scale structure (e.g. Cooray \& Sheth 2002, and
references therein), it has been shown that the evolved mass field
is approximately spatially localized if the initial perturbations
are Gaussian. Although gravitational coupling is long-term, the
spatial locality in the phase space is not disturbed by the
non-linear evolution of cosmic mass field. This property
essentially is due to the self-similarity of the hierarchical
clustering. This result has been tested with high resolution
N-body simulation samples (Feng \& Fang 2004).

In this paper, we investigate the spatial locality of phase space
correlations with real sample -- the galaxies selected from the 2MASS
Extended Source Catalog
(XSC, Jarrett et al. 2000). Our motivation is two-fold. First, the spatial
locality provides a test of the Gaussianity of the initial density
perturbations on small scales. Although many tests on the Gaussianity of
the initial perturbations have been done with the temperature fluctuations
of the Cosmic Microwave Background Radiation (e.g. Komatsu, et al. 2003,
Pando, Valls-Gabaud \& Fang, 1998), the comoving scales of these tests
are not less than a few Mpc. The spatial locality of the 2MASS samples can
test the Gaussianity to scale as small as about 0.04 h$^{-1}$ Mpc. Second,
the distribution of galaxies is biased from the mass field of dark matter.
Some bias models assume that the relation between the distribution of
galaxies and underlying mass field is stochastic and nonlocal (e.g.
Dekel \& Lahav, 1999). This mechanism will lead to nonlocality
of the galaxy correlation function in phase space, even when the underlying
dark matter mass field is spatially localized. Therefore, the spatial
locality should be effective in testing the stochasticity and nonlocality
of bias models.

The outline of this paper is as follows. \S 2 presents the
statistics and dynamics of the spatial locality in phase space. \S
3 describes the basic properties of the 2MASS galaxy samples with
a space-scale decomposition. The results of the spatial locality
of the 2MASS galaxy distribution are presented in \S 4. Finally,
the conclusions and discussions are be given in \S 5.

\section{Statistics and dynamics of spatial locality in phase space}

\subsection{Variables of the mass field in phase space}

The cosmic mass density contrast field $\delta$ in ${\bf x}$-space is given
by $\delta({\bf x})=\langle {\bf x}|\delta\rangle$, in ${\bf k}$-space by
$\hat{\delta}({\bf k})=\langle {\bf k}|\delta\rangle$. These
two descriptions are equivalent, and the set of the bases $|{\bf x}\rangle$
or $|{\bf k}\rangle$ are complete and orthogonal. $\delta({\bf x})$ and
$\hat{\delta}({\bf k})$ correspond to, respectively,
the coordinate- and momentum-representations of quantum mechanical
wavefunctions. To study the statistical and dynamical behavior of the mass
field in $k-x$ phase space, we should describe the field $\delta$ in the
Wigner representation. Therefore, the phase space
analysis of $\delta$ can be performed with the discrete wavelet transform
(DWT). The Wigner wavefunction is a prototype of the DWT.

For the DWT analysis, we use the same notation as Fang \& Feng
(2000) or Fang \& Thews (1998). In the DWT scheme, there are two
sets of base given by 1.) scaling functions $\phi_{\bf j,l}({\bf
x})=\langle {\bf x}|{\bf j,l}\rangle_s$ and 2.) wavelets
$\psi_{j,l}({\bf x})=\langle {\bf x}|{\bf j,l}\rangle_w$, where
$j=0,1,..$ and $l=0, 1...2^{j}-1$. In 1-D space with size $L$, the
scaling function $\phi_{ j,l}( x)$ is localized in physical space
$lL/2^{j}  < x \leq (l+1)L/2^{j}$, while wavelet $\psi_{j,l}$ is
localized in phase space cell $lL/2^{j}  < x \leq (l+1)L/2^{j}$
and $\pi 2^{j}/L < |k| <(3/2)2\pi 2^{j}/L$. The DWT base in 3-D
space is given by a direct product of the 1-D base, i.e. $|{\bf
j,l}\rangle_s=|j_1,l_1\rangle_s |j_2,l_2\rangle_s
|j_3,l_3\rangle_s$ and $|{\bf j,l}\rangle_w=|j_1,l_1\rangle_w
|j_2,l_2\rangle_w |j_3,l_3\rangle_w$. If a sample is in a cubic
box of $0\leq x_i \leq L$, $i=1,2,3$ and volume $V=L^3$, the bases
$|{\bf j,l}\rangle_w$ are localized in the spatial range
$l_iL/2^{j_i}  < x_i \leq (l_i+1)L/2^{j_i}$ and wavenumber range
$\pi 2^{j_i}/L < |k_i| <(3/2)2\pi 2^{j_i}/L$. The simplest scaling
and wavelet are given by
\begin{eqnarray}
\phi_{j,l}(x)= \langle x|j,l\rangle_s & = & \left\{\begin{array}{ll}
                        \sqrt{2^j/L} & \mbox{if $lL/2^j< x< (l+1)L/2^j$}\\
                        0 & \mbox{otherwise}
                    \end{array}
    \right .
 \nonumber \\
\psi_{j,l}(x)=\langle x|j,l\rangle_w & = & \left\{\begin{array}{ll}
            \sqrt{2^j/L} & \mbox{if $lL/2^j< x< [l+1/2]L/2^{j}$}\\
           -\sqrt{2^j/L} & \mbox{if $[l+1/2]L/2^{j}< x< (l+1)L/2^{j}$}\\
                      0 & \mbox{otherwise}
                    \end{array}
    \right . \nonumber
\end{eqnarray}
This is the so-called Haar wavelet. However, we will not use the Haar
wavelet in the numerical calculation of \S 3 and 4, because the discontinuity
of the Haar function made it not to be very well localized in the
Fourier space ($k$-space). We will apply the wavelet of Daubechies 4
(Daubechies, 1992), which has much better behavior in $k$ space. The
scaling and wavelet of the Daubechies 4 in $x$- and $k$-spaces can be
found in the Numerical Recipes (Press et al 1992) and Yang et al (2001).
The properties of the scaling function and wavelet discussed in this
section are generally available for either the Haar, Daubechies or
other DWTs.

The scaling functions are orthogonormal with respect to index ${\bf l}$ as
\begin{equation}
\ _{s}\langle {\bf j,l}| {\bf j,l'}\rangle_{s}
\equiv\int \phi_{\bf j,l}({\bf x})\phi_{\bf j,l'}({\bf x})d{\bf x}
 = \delta^K_{\bf l,l'}.
\end{equation}
Scaling function $\phi_{\bf j,l}({\bf x})$ actually is a window function
for the spatial range $l_iL/2^{j_i}  < x_i \leq (l_i+1)L/2^{j_i}$. The
scaling function coefficient (SFC) of a density field $\rho$ is defined by
$\epsilon_{\bf j,l}\equiv \ _{s}\langle {\bf j,l}| \rho \rangle
=\int \rho({\bf x})\phi_{\bf j,l}({\bf x})d{\bf x}$, and
therefore, the mean density in 3-D spatial range of
$l_iL/2^{j_i}  < x_i \leq (l_i+1)L/2^{j_i}$ is
\begin{equation}
\rho_{\bf j,l}=\frac{\epsilon_{\bf j,l} }
 {\ _{s}\langle {\bf j,l}| 1 \rangle },
\end{equation}
where $| 1 \rangle$ is a uniform field with density equal to unity. The
density field can be expressed by the SFCs as
\begin{equation}
\rho({\bf x})=\rho^{(\bf j)}({\bf x})+  O(>{\bf j}),
\end{equation}
where  $l_i$ runs 0, 1,...$2^{j_i}-1$ and $O(>{\bf j})$ means all
fluctuations on scales less than $(L^3/2^{j_1+j_2+j_3})^{1/3}$, and
\begin{equation}
\rho^{(\bf j)}({\bf x})= \sum_{\bf l}
\epsilon_{\bf j,l}\phi_{\bf j,l}({\bf x}).
\end{equation}
Therefore, $\rho^{(\bf j)}({\bf x})$ is the density field smoothed on scale
$(L^3/2^{j_1+j_2+j_3})^{1/3}$. The SFCs $\epsilon_{\bf j,l}$ and
$\rho_{\bf j,l}$ are similar to the mass field variables given by
count-in-cell. Thus, the two-mode correlation
$\langle \rho_{\bf j,l}\rho_{\bf j,l'}\rangle$ will display the similar
features as ordinary two-point correlation function, which generally is
not localized for Gaussian fields with a $k$-dependent Fourier
power spectrum $P(k)$.

The wavelets $\psi_{\bf j,l}({\bf x})$ are orthogonal with respect to
both indexes ${\bf j}$ and ${\bf l}$
\begin{equation}
\ _s\langle {\bf j,l}|{\bf  j',l'}\rangle_s \equiv
\int \psi_{\bf j,l}({\bf x})\psi_{\bf j',l'}({\bf x})d{\bf x}
 =\delta_{\bf j,j'}\delta_{\bf l,l'}.
\end{equation}
The wavelet function coefficient (WFC) of the density contrast field
$\delta$ is defined by
\begin{equation}
\tilde{\epsilon}_{\bf j,l}\equiv \ _{w}\langle {\bf j,l}| \delta \rangle
=\int \delta({\bf x})\psi_{\bf j,l}({\bf x})d{\bf x}.
\end{equation}
Since the set of the wavelet bases $|{\bf j,l}\rangle_w$ is complete,
the density contrast field can be expressed as
\begin{equation}
\delta({\bf x})=\sum_{\bf j}\sum_{\bf l}\tilde{\epsilon}_{\bf j,l}
\psi_{\bf j,l}({\bf x}),
\end{equation}
where each $j_i$ runs 0, 1, 2... and $l_i$ runs 0,
1,...$2^{j_i}-1$. Therefore, the WFCs $\tilde{\epsilon}_{\bf j,l}$
can be used as the variables of the mass field $\delta$.
$\tilde{\epsilon}_{\bf j,l}$ is the fluctuations of the density
field around scales ${\bf k}=2\pi {\bf n}/L$ with ${\bf
n}=(2^{j_1},2^{j_2},2^{j_3})$ and at the physical area ${\bf l}$
with size $\Delta {\bf x}=(L/2^{j_1},L/2^{j_2},L/2^{j_3})$. In
each dimension $i$, $\tilde{\epsilon}_{\bf j,l}$ is given by a
superposition of fluctuations in the waveband $k_i \pm \Delta
k_i/2$, where $k_i=2\pi 2^{j_i}/L$, and $\Delta k_i=2\pi/\Delta
x_i = k_i$. Therefore, $\tilde{\epsilon}_{\bf j,l}$ can play the
role as the variables $\delta({\bf k,x})$ of eq.(2). We have then
\begin{equation}
\langle \tilde{\epsilon}_{\bf j,l}
   \tilde{\epsilon}_{\bf j',l'}\rangle
  = P_{\bf j} \delta^K_{\bf j,j'}\delta^K_{\bf l,l'}.
\end{equation}
As we have emphasized, the locality of the factors $\delta^K_{\bf j,j'}$ and
$\delta^K_{\bf l,l'}$ in eq.(10) generally are very good approximation
for Gaussian field. The factor $P_{\bf j}$ in eq.(10) is the DWT power
spectrum of the mass density perturbations (Fang \& Feng 2000).
$P_{\bf j}$ is not dependent on ${\bf l}$, as cosmic mass field is
assumed to be randomly uniform.

The spatial locality $\delta^K_{\bf l,l'}$ arises from the phase
decoherence given by the superposition of the Fourier
modes in the band ${\bf k}$ to ${\bf k+\Delta k}$. This point can easily
be seen from the orthogonal relation eq.(7) by taking ${\bf j=j'}$.
For the 1-D case we have
\begin{eqnarray}
\delta^K_{l,l'} & = & \int \psi_{j,l}(x)\psi_{j,l'}(x)dx \\ \nonumber
  &  = &\frac{1}{L}\sum_{n=-\infty}^{\infty}\hat{\psi}_{j,l}(k)
     \hat{\psi}_{j,l'}(k) \\ \nonumber
  & = &
     \frac{1}{2^j}\sum_{n=-\infty}^{\infty}\hat{\psi}^2(n/2^j)
         e^{-i2\pi n(l-l')/2^j},
\end{eqnarray}
where $k=2\pi n/(L/2^j)$, and
$\hat{\psi}_{j,l}(k)$ is the Fourier
counterpart of $\psi_{j,l}(x)$. The last step of deriving eq.(11) used the
following relation (Fang \& Thews 1998)
\begin{equation}
\hat{\psi}_{j,l}(k)= \left (\frac{L}{2^j}\right )^{1/2}
  \hat{\psi}(n/2^j)e^{-i2\pi nl/2^j},
\end{equation}
where $\hat{\psi}(z)$ is the Fourier transform of the basic wavelet.
It is non-zero only within $z= 1 \pm 1/2$ and  $-1 \pm 1/2$.
Therefore $\hat{\psi}(n/2^j)$ is non-zero only for
$2^j(1-1/2)<|n|<2^j(1+1/2)$. Since the factor $\hat{\psi}^2(n/2^j)$
is always positive, the spatial locality factor $\delta^K_{l,l'}$ of
eq.(11) basically comes from the average of the phase factor
$e^{-i2\pi n(l-l')/2^j}$ over $n$. For $(l-l')\neq 0$, the phase
$2\pi n(l-l')/2^j$ is uniform distributed in range
$(1/2)2\pi - (3/2)2\pi$. Therefore, the phase is similar to a randomly
uniform distributed. Thus, for a function $f(x)$, we have approximately
\begin{equation}
\int \psi_{ j,l}(x)f(x)\psi_{j,l'}( x)d x
   = \frac{1}{2^j}\sum_{n=-\infty}^{\infty}\hat{f}(k)\hat{\psi}^2(n/2^j)
         e^{-i2\pi n(l-l')/2^j} \propto  \delta^K_{l,l'},
\end{equation}
if $\hat{f}(k)$, the Fourier counterpart of $f(x)$, does not change
the average of the phase factor $e^{-i2\pi n(l-l')/2^j}$ in the
range $2^j(1-1/2)<|n|<2^j(1+1/2)$.

\subsection{Statistical criterion's for spatial locality}

The mass field evolution under a self-similarly hierarchical process
will not violate the initial spatial locality (Feng \& Fang 2004). The
correlations between DWT modes at different physical positions
are always substantially less than that at the same position. In other
words, gravitational evolution causes a strong coupling between the phase
space modes $({\bf j,l})$ and $({\bf j',l'})$ with different scales
${\bf j}$ and ${\bf j'}$, but with spatial range of ${\bf l}$ and
${\bf l'}$ overlapped. The coupling between modes at different
physical range is very weak. The factor $\delta^K_{\bf l,l'}$ in the
initial Gaussian field [eq.(10)] is memorized during the nonlinear
evolution.

The spatial locality can be measured by
\begin{equation}
\kappa_{\bf j} \equiv
\frac{\langle \tilde{\epsilon}_{\bf j,l}\tilde{\epsilon}_{\bf j,l'}\rangle}
{\langle |\tilde{\epsilon}_{\bf j,l}|^2\rangle^{1/2}
 \langle |\tilde{\epsilon}_{\bf j,l'}|^2\rangle^{1/2}}
= \left \{ \begin{array}{ll}
           1      &  {\bf l=l'} \\
          \ll 1 & {\bf l \neq l'}\\
   \end{array} \right .
\end{equation}
where $\langle ... \rangle$ is average over the ensemble of variables
$\tilde{\epsilon}_{\bf j,l}$ with $l_i=0...2^{l_i}-1$ for a given
${\bf j}$. Considering the mass field $\rho$ is randomly uniform, we have
$\langle |\tilde{\epsilon}_{\bf j,l}|^2\rangle=
\langle |\tilde{\epsilon}_{\bf j,l'}|^2\rangle$. Therefore, eq.(14) can
also be rewritten as
\begin{equation}
  \frac {|\int d{\bf x} \int d{\bf x'}
     \psi_{\bf j,l}({\bf x})\xi( {\bf x- x'})
     \psi_{\bf j,l'}({\bf x'})|}
   {|\int d{\bf x} \int d{\bf x'}
     \psi_{\bf j,l}({\bf x})\xi( {\bf x- x'})
     \psi_{\bf j,l}({\bf x'})|
} \ll 1, \hspace{5mm} {\rm if \ \ } {\bf l}\neq {\bf l'}.
\end{equation}
where $\xi( {\bf x- x'})$ is the two-point correlation function of
mass field $\rho$. Eq.(15) is a criterion of the spatial locality
with second order statistics.

One can construct the spatial locality with higher order
correlations of the $\tilde{\epsilon}_{\bf j,l}$. For a $(p+q)$
order statistical criterion, we use statistics as
\begin{equation}
C^{p,q}_{\bf j} \equiv
\frac{\langle \tilde{\epsilon}^p_{\bf j,l}
  \tilde{\epsilon}^q_{\bf j,l'}\rangle}
{\langle \tilde{\epsilon}^{p}_{\bf j,l}\rangle
 \langle \tilde{\epsilon}^{q}_{\bf j,l'}\rangle},
\end{equation}
where $p$ and $q$ can be any even number. Obviously, for Gaussian
fields. $\langle \tilde{\epsilon}^p_{\bf
j,l}\tilde{\epsilon}^q_{\bf j,l'}\rangle=\langle
\tilde{\epsilon}^p_{\bf j,l}\rangle \langle
  \tilde{\epsilon}^q_{\bf j,l'}\rangle$ at ${\bf l \neq l'}$, and
therefore, $C_{\bf j}^{p,q}=1$ at ${\bf l \neq l'}$. If
$C^{p,q}_{\bf j} \neq 1$ at ${\bf l\neq l'}$, the field is nonlocally
correlated. Therefore, the spatial local evolution is given by
\begin{equation}
C^{p,q}_{\bf j} \simeq 1 {\rm \ \ if \ \ }  {\bf l\neq l'}.
\end{equation}
Eqs.(14) and (17) are the basic statistical criterion's used for
testing the spatial locality.

\subsection{Bias model of galaxies and spatial locality of correlation}

The number density of galaxies $\rho_g({\bf x})$ are not simply
proportional to the mass density $\rho({\bf x})$ of dark matter.
It is biased. The simplest linear model of galaxies bias is given
by $\delta_g({\bf x})= b\delta({\bf x})$, where $\delta_g({\bf
x})$ is the number density contrast of galaxies, and $b$ the bias
parameter. This model implies that the relation between
$\delta_g({\bf x})$ and $\delta({\bf x})$ is deterministic and
localized. It has been argued that the locality assumption is not
trivial, and  consequently, stochastic and nonlinear bias models have
been proposed. In these models, the galaxy field $\delta_g({\bf
x})$ at ${\bf x}$ is not determined only by dark
matter field $\delta({\bf x})$ at the same point ${\bf x}$. The
relation between $\delta_g({\bf x})$ and $\delta({\bf x})$ is
stochastic and non-local. In these models, the two-point
correlation function of galaxies $\xi_g({\bf x-x'})$ is related to
the two-point correlation function $\xi({\bf x-x'})$ of dark
matter by (Matsubara 1999)
\begin{equation}
\xi_g({\bf x-x'})=
  \int K({\bf y})K({\bf y'})\xi({\bf x - x' + y - y'})d{\bf y}d{\bf y'}
   + {\rm higher\  order\ terms},
\end{equation}
where $K({\bf y})$ describes the nonlocal nature of the bias. A
galaxy at ${\bf x}$ is dependent on the dark matter field
$\delta({\bf x+y})$ with probability proportional to $K({\bf y})$.
If $K({\bf y})$ is non-zero in the range $|{\bf y}|< D$, the
relation between  galaxies and dark matter field is nonlocal with
size $D$. Thus, the galaxy correlation of phase space modes
becomes
\begin{eqnarray}
\langle \tilde{\epsilon}_{\bf j,l}\tilde{\epsilon}_{\bf
j,l'}\rangle & = & \int \psi_{j,l}({\bf x})\xi_g({\bf x-x'})
\psi_{j,l'}({\bf x'})
  d{\bf x}d{\bf x'} \\ \nonumber
 & = & \frac{1}{2^{j_1+j_2+j_3}}\sum_{n_1,n_2,n_3=-\infty}^{\infty}
  \hat{K}({\bf k})P({\bf k})\hat{\psi}^2(n_1/2^{j_1})\hat{\psi}^2(n_2/2^{j_1})
   \\ \nonumber
  & &    e^{-i2\pi n_1(l_1-l'_1)/2^{j_1}}
         e^{-i2\pi n_2(l_2-l'_2)/2^{j_2}}e^{-i2\pi n_1(l_3-l'_3)/2^{j_3}},
\end{eqnarray}
where $\hat{K}({\bf k})$ is the Fourier counterpart of $K({\bf
y})$. $\hat{K}({\bf k})$ is non-zero only in the band $|{\bf k}| <
2\pi/D$.

Thus, if $D$ is large than $L/2^j$, the non-zero range of
$\hat{K}({\bf k})$ is limited in $n_i <L 2^{j_i}/D$ or
$n_i/2^{j_i}<1$. In this narrow range of $n_i$, the average of the
phase factor $e^{-i2\pi n_1(l_i-l'_i)/2^{j_i}}$ cannot be zero
when $|l_i-l'_i|\geq 1$. Therefore, even when the underlying field
of dark matter is spatially localized as eq.(15), the galaxy
clustering with nonlocal bias will not be spatially localized.
Hence, if the spatial locality holds for a given scale $j$, the
nonlocal size of galaxy bias should not be larger than $L/2^j$,
i.e. the nonlocal scale $D$ will not be larger than that given by
the uncertainty relation $\Delta x = 2\pi/k$.

\subsection{2-D distributions}

Our sample of the 2MASS galaxies is 2-D. With the DWT, the
projection of 3-D field $\delta(x_1,x_2,x_3)$ to 2-D
$\delta(x_1,x_2)$ is given by
\begin{equation}
\delta(x_1,x_2)  =\ _s\langle j_3,0|\delta\rangle
   =  \int \delta(x_1,x_2,x_3)\phi_{j_3,0}(x_3)dx_3,
\end{equation}
where $x_1$ and $x_2$ are coordinates on sky plane, while $x_3$
stands for the spatial coordinate in redshift direction. In
eq.(20), $\phi_{j_3,0}(x_3)$ is a scaling function on scale
$L/2^{j_3}$. Hence, if $L/2^{j_3}$ is equal to the depth of the
sample in the redshift direction, eq.(20) is a projection of a 3-D
sample $\delta(x_1,x_2,x_3)$ with depth $L/2^{j_3}$ onto a 2-D
distribution. The DWT variables in 2-D analysis is then
\begin{eqnarray}
\tilde{\epsilon}_{\bf j,l}^{2D} & = &
  \int \delta(x_1,x_2)\psi_{j_1,l_1}(x_1)\psi_{j_2,l_2}(x_2)dx_1dx_2
   \\ \nonumber
 & = &
\int \delta(x_1,x_2,x_3)
\psi_{j_1,l_1}(x_1)\psi_{j_2,l_2}(x_2)\phi_{j_3,0}(x_3),
dx_1dx_2dx_3.
\end{eqnarray}
Eq.(21) shows $\tilde{\epsilon}_{\bf j,l}^{2D}$ is a sampling of
$\delta({\bf x})$ by basis
$|j_1,l_1\rangle_w|j_2,l_2\rangle_w|j_3,0\rangle_s$, which is
mixed from wavelets $|j_1,l_1 \rangle_w$,
 $|j_2,l_2 \rangle_w$ and scaling
$|j_3,0 \rangle_s$.

Since the $|j_1,l_1\rangle_w$ and $|j_2,l_2\rangle_w$ are
orthogonal with $|j_3,0\rangle_s$, the DWT is effective in
distinguishing the redshift direction. One can repeat the analysis
of Feng \& Fang (2004) to show that the correlation of the
variables $\tilde{\epsilon}_{\bf j,l}^{2D}$ will be spatially
localized with respect to the position indexes $l_1$ or $l_2$
regardless the projection of the redshift direction. The reason is
simple. The scaling function base $|j_3,0\rangle_s$ counts all
galaxies in the redshift direction. However, the wavelet bases
$|j_1,l_1\rangle_s$ and $|j_2,l_2\rangle_s$ measure the {\it
difference} between the densities of two neighboring cells on
scale $j_1+1$ and $j_2+1$, and therefore, the background of the
two neighbor cells are cancelled. The background may cause
uncertainty of the shot noise, but it does not contribute to the
difference $\tilde{\epsilon}_{\bf j,l}^{2D}$. Thus, if a 3-D mass
field is spatially localized, the 2-D variables
$\tilde{\epsilon}_{\bf j,l}^{2D}$ should satisfy eqs.(14) and
(17). We do not write the superscript $2D$ below to simplify the
notation.

\section{Samples for the spatial locality analysis}

\subsection{Galaxies selected from 2MASS-XSC}

The galaxy samples used for the spatial locality analysis are
selected from the 2MASS extended source catalog (XSC, Jarrett et
al. 2000), which covers almost the entire sky at wavelength
between 1 and 2 $\mu$m. To select galaxies, we use the indicator
${\rm K\_m\_k20fe}$, which measures the magnitude inside a
elliptical isophote with surface brightness of 20 mag ${\rm
arcsec^{-2}}$ in $K_s$-band (from then on, we infer ${\rm
K\_m\_k20fe}$ as $K_s$). There are approximately 1.6 million
extended objects with $K_s<14.3$. Most of the XSC sources at $|b|
> 20^{\circ}$ are galaxies ($>98\%$). The contamination mainly is
from stars. The reliability of separating stars from extended
sources is $95\%$ at $|b| > 10^{\circ}$, but drops rapidly to $<
65\%$ at $|b| > 5^{\circ}$. To avoid this contaminant, we use a
latitude cut of $|b| > 10^{\circ}$. We also removed a small number
of bright ($K_s<9$) sources  by the parameters of the XSC
confusion flag (${\rm cc\_flag}$) and visual verification score
for source (${\rm vc}$). They are identified as non-extended
sources including artifacts. Moreover, to eliminate duplicate
sources and have a uniform sample, we use the parameters ${\rm
use\_src=1}$ and ${\rm dup\_src=0}$ \footnote{The notations of the
2MASS parameters used in this paragraph are from the list shown in
the 2MASS Web site
http://www.ipac.caltech.edu/2mass/releases/allsky/doc.}.

To select the range of $K_s$, we use the standard $\log N-\log S$
test to examine the completeness of the sample. The number counts can
be approximated by a power-law as
\begin{equation}
\frac{dN}{dm} \propto 10^{\,\kappa\,m}.
\end{equation}
The XSC sources with $|b| > 30^{\circ}$ and $12<K_s<13.7$ are believed
to be galaxies with 99\% reliability (Maller et al. 2003).
For this sample, the index $\kappa$ is found to be $0.641 \pm 0.006$.
If considering this $\kappa$ to be the standard, the completeness
of a sample can be estimated by the deviation of $(dN/dm)_{sample}$ from
the standard (Afshordi, Loh \& Strauss 2003), i.e.
\begin{equation}
C(m)= \frac{(dN/dm)_{standard}}{(dN/dm)_{sample}}.
\end{equation}

Figure 1 shows the number counts $dN/dm$ and completeness $C(m)$ for
sample (1) (the standard) of galaxies with $|b| > 30^{\circ}$ and
$12<K_s<13.7$, and sample (2) with $|b| > 10^{\circ}$ region when
$9<K_s<14$. The figure shows that in the range
$11<K_s<13.7$, sample (2) has the same number counts and completeness
as sample (1). The completeness $C(m)$ of the sample (2) is obviously
larger than sample (1) when $K_s<10.0$. This indicates the catalog to be
contaminated towards the bright end. $C(m)$ drops below 0.9 when $K_s>14.0$.
Thus we use a cut of $10.0<K_s<14.0$ to ensure our sample to be complete
greater than $90\%$. This sample contains 987,125 galaxies.

To carry out a 2-D DWT analysis, we first to take an equal-area projection
with the Lambert azimuthal algorithm:
\begin{eqnarray}
x_1~=~R\sqrt{2-2|\sin\textit{b}|}\cos\textit{l}, \\ \nonumber
x_2~=~R\sqrt{2-2|\sin\textit{b}|}\sin\textit{l},
\end{eqnarray}
where R is a relative scale factor, $\textit{b}$ is the Galactic
latitude and $\textit{l}$ is the Galactic longitude. This
hemisphere scheme projects the whole sky into two circular plane,
northern and southern sky. From each circular plane, we select 14
squares, each of which has an angular size of about
$28.28^{\circ}\times28.28^{\circ}$ (800 square degrees). The
angular size labelled by $j$ is $28.28/2^{j}$. These squares do
not overlap each other to guarantee the independence of
statistics. Each square contains $\sim$ 28,000 to 41,000 galaxies,
and 35,000 in average. The 28 ($28.28^{\circ}\times28.28^{\circ}$)
squares are our samples for statistics. Since XSC galaxies are
resolved to 10$^{''}$, our analysis can reach to angular scale of
0.01 degree. For each square, we also produce random samples by
randomizing the coordinate $(x_1,x_2)$ of galaxies.

\subsection{Two-Point correlation functions}

The 2-D number density distribution of $N$ galaxies with coordinate
$(x^n_1, x^n_2)$ ($n=1...N$) is given by
\begin{equation}
\rho({\bf x})=\sum_n \delta^D({\bf x-x}^n)
\end{equation}
where $\delta^D$ is the Dirac delta function. We first calculate
the correlation between modes of the density variable eq.(4),
$\langle \rho_{\bf j,l} \rho_{\bf j,l'}\rangle$. The variable
$\rho_{\bf j,l}$ is the mean number density in a 2-D spatial range
$l_iL/2^{j_i}  < x_i \leq (l_i+1)L/2^{j_i}$, $i=1,2$. As have been
discussed in \S 2.1, the correlation function $\langle \rho_{\bf
j,l} \rho_{\bf j,l'}\rangle$ is similar to the ordinary two-point
angular correlation function $w(\theta)$, where $\theta$ is
angular distance. A difference between $w(\theta)$ and $\langle
\rho_{\bf j,l} \rho_{\bf j,l'}\rangle$ is that $w(\theta)$ is
calculated by counting the pair within a given angular distance,
while the DWT algorithm is based on variable $\rho_{\bf j,l}$
corresponding to cell $({\bf j,l})$ in phase space. It has been
pointed out that the method of counting pairs actually is
pair-weighted (Strauss, Ostriker \& Cen 1998). That is, a cell
$({\bf j,l})$ at the dense regions will be counted more than one
time in the pair-weighted statistics, while empty cells contribute
nothing. On the other hand, in the DWT analysis, each phase space
cell $(\bf j,l)$ supports only {\it one} variable $\rho_{\bf j,l}$
regardless the cell is dense, or empty of galaxies.

Figure 2 presents the correlation function $\langle \rho_{\bf j,l}
\rho_{\bf j,l'}\rangle$ for mode ${\bf j}=(j,j)$ and $j=8$, 9, 10
and 11, corresponding, respectively, to smooth the sample on
angular scale $28.3/2^j$ in unit of angular degrees. The angular
distance in Fig. 2 is given by $\theta=28.3|{\bf l-l'}|/2^j $. The
error bar is one sigma among the samples of 28 correlation functions
given by the 28 data squares. As expected that the $\theta$-dependencies
of the correlation function
show the typical power law in the angular range from $\simeq$ 0.01 to
1 degree. For $j= 8$, 9, the power law breaks at $\simeq$ 1 degree,
while for $j=11$ correlation function is of a power law till scale
$\simeq$ 2 degree. The best fitting results for a power law
$\langle \rho_{\bf j,l} \rho_{\bf j,l'}\rangle= A\theta^{-\beta}$ are
shown by the solid lines in Fig. 2. The amplitude $A$ and index $\beta$ are
listed in Table 1.

\begin{table}[t]
\caption{Power law fitting of correlation function
   $\langle\rho_{\bf j,l} \rho_{\bf j,l'}\rangle$}
\bigskip
\begin{tabular}{ccc}
\tableline
  j & amplitude $A$ &  index $\beta$ \\ \tableline
  8 & 0.154 $\pm$ 0.063   & 1.020 $\pm$   0.033    \\
  9 & 0.183 $\pm$ 0.002   & 0.845 $\pm$   0.017 \\
  10& 0.190 $\pm$ 0.002   & 0.819 $\pm$   0.012 \\
  11 & 0.200 $\pm$ 0.004  & 0.800 $\pm$  0.008 \\
\tableline
\end{tabular}
\end{table}

Table 1 shows a systematic decrease of the index $\beta$ with the
increase of $j$. Since the scaling functions are low pass filters, the
SFCs at $j$ smooth out all fluctuations on angular scales less than
$28.3/2^j$. Therefore, the $j$-dependence of $\beta$ indicates that
the clustering of the 2MASS galaxies is stronger (and $\beta$ is smaller)
at smaller scales (or larger $j$). This result is consistent with the general
picture that smaller scale clustering enters the nonlinear regime earlier.
The density field of the 2MASS galaxies on small scales has significantly
undergone nonlinear evolution. The values of $\beta$ in Table 1 are little
larger than that given by the ordinary two-point angular correlation function
(e.g. Maller et al 2003). This is due probably to the fact that the
pair-weighted algorithm gives a high weight to dense regions, and
therefore, high weight to small scales.

\section{Spatial locality of the 2MASS galaxies}

\subsection{Second order correlations}

We now turn to the major task of this paper -- to
check whether the number density field of the 2MASS galaxies is
spatially localized. We start by looking at $\kappa_{\bf j}$ as given by
eq.(14). The result of $\kappa_j$ vs. $\theta$ is shown in Fig. 3.
Here we consider only modes with ${\bf j}=(j,j)$ and $j=6$ - 10. The
error bars are estimated by 90\% confidence level for all 28 data squares.

Fig. 3 shows that the WFC correlation function
$\langle\tilde{\epsilon}_{\bf j,l} \tilde{\epsilon}_{\bf j,l'}\rangle$
is substantially different from
$\langle \rho_{\bf j,l} \rho_{\bf j,l'}\rangle$. The WFC
correlation function is spatially localized for all cases
considered. All nonlocal ($|{\bf l-l'}| \geq 1$) correlations are much
less than 1, and practically equal to zero within the error bars. In other
words, the correlation length in terms of the position index ${\bf l}$
is zero.

The spatial locality is not a surprise on large scales for which
the field basically remains Gaussian. However, Fig. 3 shows
that spatial locality holds on scales as small as
$\theta=28.2/2^{10} \simeq 0.03$ degrees. This result is
in contrast with Fig. 2, in which the correlation function
$\langle \rho_{\bf j,l} \rho_{\bf j,l'}\rangle$ at $|{\bf l-l'}|=
1$ for all $j$ is always comparable with that of $|{\bf l-l'}|= 0$.
Therefore, it is appropriate to describe the correlation function by
$\langle\tilde{\epsilon}_{\bf j,l}\tilde{\epsilon}_{\bf j',l'}\rangle
 \propto \delta^K_{\bf l,l'}$.

One cannot conclude that the spatial locality is due to the
dynamical evolution because spatial locality can also be given by
the Gaussianization due to the central limit theorem (CLT).
According to the CLT, a variable given by a superposition of many
independent identical random variables tends to become Gaussian.
If we consider that a 2-D field is a superposition of many
identical slides of a 3-D field, the 2-D field variables will be
Gaussianized. If so, the spatial locality is a trivial result.
However, the 2-D WFC $\tilde{\epsilon}_{\bf j,l}$ are not subject
to  the CLT. This is because, as discussed in \S 2.3
$\tilde{\epsilon}_{\bf j,l}$ is not a 3-D superposition. This
point can be directly shown with the non-Gaussianity of variable
$\tilde{\epsilon}_{\bf j,l}$. We plot in Figure 4 the one-point
distributions of the ensemble of $\tilde{\epsilon}_{\bf j,l}$ for
${\bf j}=(j,j)$ with  $j=5$ - 10 and $l_i=1...2^j-1$. All the
distributions of Fig. 4 are non-Gaussian. These one-point
distributions generally are long tailed, and therefore, the
kurtosis of this field is high.

We also see from Fig. 4 that the non-Gaussianity of the one-point
distributions is scale-dependent. On
large scales $j=5$ and 6, the distribution consists of a Gaussian
center and a power law tail. On scale $j=7$, the Gaussian center
disappears, and the whole distribution is a power law. On small
scales $j\geq 8$, the distributions still have a power law tail, but
the sharp center part appears and grows with $j$. This means that in
most places the perturbations are small (center part of the one-point
distributions), but there are rare events with very large perturbations
(long tail). That is, there is no Gaussianization of the variable
$\tilde{\epsilon}_{\bf j,l}$. The spatial locality doesn't arise from
the Gaussianization of $\tilde{\epsilon}_{\bf j,l}$.

In Figure 5, we plot the one-point distributions of the SFCs
$\epsilon_{\bf j,l}$. These distributions are non-Gaussian too. They
have a long-tail towards high density, or high value of
$\epsilon_{\bf j,l}$. These distributions are similar to the
one-point distributions given by count-in-cell. This result shows
again that both the WFCs and SFCs are highly non-Gaussian. The spatial
locality cannot be due to the Gaussianization, but a remain
of the spatial locality of the initial perturbations.

\subsection{Fourth order correlations}

Before looking at the spatial locality of the WFC correlation at higher order
 we first calculate a 4$^{th}$ order correlation of the SFC
$\epsilon_{\bf j,l}$ defined by
\begin{equation}
S^{2,2}_{\bf j} =
\frac{\langle \epsilon^2_{\bf j,l}
  \epsilon^2_{\bf j,l+\Delta l}\rangle}
{\langle \epsilon^{2}_{\bf j,l}\rangle
 \langle \epsilon^{2}_{\bf j,l+\Delta l}\rangle}.
\end{equation}
Figure 6 plots the $j$-dependence of $S^{2,2}_{\bf j}$ for ${\bf
j}=(j,j)$ with $|\Delta{\bf l}|=1$, 2 and 3 for both the 2MASS and
random samples. The error bars are again the 90\% confidence of
the 28 samples. As expected the random samples always show
$S^{2,2}_{\bf j} =1$, i.e. there is no correlation. On small
scales $j=10$, $S^{2,2}_{\bf j}$ for random sample is little
larger than 1 at $\Delta=1$. This is because of shot noise, i.e.
the mean number per cell on small scales is very small. On the
other hand, the 2MASS galaxy samples show significant non-local
correlation $S^{2,2}_{\bf j}  > 1$ for all scales $j > 5$. That
is, the correlation length for 4th order statistics of the SFC
variables is non-zero. The correlation is no localized. On large
scales $j=2$ - 4, the nonlocal correlation of the 2MASS galaxy
samples is negligible, because the field is still in Gaussian on
such large scales.

However, the 4$^{th}$ order correlation function of the
$\tilde{\epsilon}_{\bf j,l}$ has completely different behavior from
the SFC. Similar to eq.(26), we consider a 4th order correlations
of the WFCs by
\begin{equation}
C^{2,2}_{\bf j} =
\frac{\langle \tilde{\epsilon}^2_{\bf j,l}
  \tilde{\epsilon}^2_{\bf j,l+\Delta l}\rangle}
{\langle \tilde{\epsilon}^{2}_{\bf j,l}\rangle
 \langle \tilde{\epsilon}^{2}_{\bf j,l+\Delta l}\rangle}.
\end{equation}
where $\Delta {\bf l} \equiv {\bf l-l'}$. Eq.(27) is
eq.(16) with $p=q=2$.  Figure 7 presents $C^{2,2}_{\bf j}$ vs.
$\theta$ for ${\bf j}=(j,j)$ with $j=6,$ 7, 8 and 9 for both the 2MASS
and random samples. The error bars in Fig. 7 are again given by the
90\% confidence levels as Fig. 6.  Figure 7 shows $C^{2,2}_{\bf j}\gg 1$ at
${\bf l=l'}=0$ for all scales considered. Nevertheless, for all non-local
cases $|\Delta {\bf l}| \geq 0$, the 2MASS samples
behave the same as the random samples within the error bars. Therefore,
the 4$^{th}$ order correlation of the WFCs is spatially localized.
Similar to the second order statistics, the spatial
locality of the 4$^{th}$ order WFC's correlation is not due to the
Gaussianization of $\tilde{\epsilon}_{\bf j,l}$. At point $|\Delta
{\bf l}|=0$, or $\theta=0$, the 2MASS samples always have much
higher value of $C^{2,2}_{\bf j}$ than that of random samples.

The difference between the WFC's and SFC's 4$^{th}$ order correlations can
be seen with the hierarchical clustering or linked pair approximation.
In this approximation, the 4$^{th}$ correlations
$\langle \tilde{\epsilon}^2_{\bf j,l}
\tilde{\epsilon}^2_{\bf j,l+\Delta l}\rangle$ can
be expressed by second order correlations as
\begin{eqnarray}
\langle \tilde{\epsilon}^2_{\bf j,l}
  \tilde{ \epsilon}^2_{\bf j,l+\Delta l}\rangle
 &  = &
\langle \tilde{\epsilon}^2_{\bf j,l}\rangle
   \tilde{\langle}\epsilon^2_{\bf j,l+\Delta l}\rangle
 +2\langle \tilde{\epsilon}_{\bf j,l}
  \tilde{\epsilon}_{\bf j,l+\Delta l}\rangle^2  \nonumber \\
 & & + Q_4 [ 2\langle \tilde{ \epsilon}^2_{\bf j,l}\rangle
\langle \tilde{\epsilon}^2_{\bf j,l+\Delta l}\rangle
\langle\tilde{\epsilon}_{\bf j,l}
   \tilde{\epsilon}_{\bf j,l+\Delta l}\rangle +
\langle \tilde{\epsilon}^2_{\bf j,l}\rangle
\langle \tilde{\epsilon}_{\bf j,l}
    \tilde{\epsilon}_{\bf j,l+\Delta l}\rangle^2 \nonumber \\
 & & + \langle \tilde{\epsilon}^2_{\bf j,l+\Delta l}\rangle
\langle \tilde{\epsilon}_{\bf j,l}
   \tilde{\epsilon}_{\bf j,l+\Delta l}\rangle^2 +
2\langle \tilde{\epsilon}_{\bf j,l}
\tilde{\epsilon}_{\bf j,l+\Delta l}\rangle^3 ],
\end{eqnarray}
where $Q_4$ is a constant. The first two terms on the r.h.s. of eq.(28) are
from unlinked diagrams and the term of $Q_4$ comes from non-linear evolution.
Obviously, if the second order correlations of the WFCs are spatially
localized [eq.(14)], the 4$^{th}$ and higher order correlations will also
be approximately spatially localized if the linked pair approximation holds.
Oppositely, the 4th order correlations will not be localized if
the second order correlations are not spatially localized.

\subsection{Scale-scale correlations of the WFCs}

In the last two subsections, we showed that the gravitational clustering
doesn't cause the WFC-WFC correlation if the two phase space modes locate
in different spatial($x$) positions. In other words, to view the
dynamical evolution of the cosmic gravitational clustering in
($k-x$) phase space, the gravitational clustering dynamics caused only
a very weak coupling in the $x$-direction. Therefore, one can expect,
the dynamical coupling should be strong in the $k$-direction, which
is coupling between modes with the same $x$, but different $k$. This is
the so-called local scale-scale coupling which we measure using
\begin{equation}
C^{2,2}_{\bf j, j+1} = \frac{\langle \tilde{\epsilon}^2_{\bf
j,{[l/2]}}
  \tilde{\epsilon}^2_{\bf j+1,l}\rangle}
{\langle \tilde{\epsilon}^{2}_{\bf j,{[l/2]}}\rangle
 \langle \tilde{\epsilon}^{2}_{\bf j+1,l}\rangle}.
\end{equation}
This measures the correlation between modes on scales ${\bf j}$ and
${\bf j+1}$. The notation [\hspace{2mm}] in Eq.(29) denotes the
integer part of the quantity since the spatial range of the cell
$(j,[l/2])$ is the same as the two cells $(j+1, l)$ and $(j+1,l+1)$.
Therefore, eq.(29) measures the scale-scale correlation at the
{\it same} physical point ${\bf x}$. It is the local scale-scale
correlation. If perturbations on scales ${\bf j}$ and ${\bf j}+1$
are uncorrelated, we have
$\langle \tilde{\epsilon}^2_{\bf j,{[l/2]}}
  \tilde{\epsilon}^2_{\bf j+1,l}\rangle =
\langle \tilde{\epsilon}^{2}_{\bf j,{[l/2]}}\rangle
 \langle \tilde{\epsilon}^{2}_{\bf j+1,l}\rangle$, and therefore,
$C^{2,2}_{\bf j, j+1}=1$.

Figure 8 presents the result of $C^{2,2}_{\bf j, j+1}$ vs. $j$ for
the 2MASS galaxies and random sample, in which ${\bf j}=(j,j)$.
This figure shows that the values of $C^{2,2}_{\bf j,j+1}$ for the
2MASS sample are larger than 1 on scales $j\geq 6$, and also
larger than random samples. Therefore, the local scale-scale
correlation is significant on scales $j\geq 6$, or $\theta \leq
0.4$ angular degree, corresponding to 1.5 h$^{-1}$ Mpc at the
median redshift of the survey. In other words, the gravitational
coupling leads to the correlation in $k$-direction, and the factor
$\delta^K_{\bf j,j'}$ in the initial perturbations eq.(10) is no
longer hold when nonlinear evolution takes places.

We can detect nonlocal scale-scale by using
\begin{equation}
C^{2,2}_{\bf j, j+1}({\bf \Delta l}) = \frac{\langle
\tilde{\epsilon}^2_{\bf j,{[l/2]}}
  \tilde{\epsilon}^2_{\bf j+1,l+\Delta l}\rangle}
{\langle \tilde{\epsilon}^{2}_{\bf j,{[l/2]}}\rangle
 \langle \tilde{\epsilon}^{2}_{\bf j,l+\Delta l}\rangle}.
\end{equation}
When $|\Delta{\bf l}| =0$, $C^{2,2}_{\bf j, j+1}(0)$ is the
same eq.(29). When $|\Delta{\bf l}|\geq 1$, $C^{2,2}_{\bf j,
j+1}(\Delta l)$ measures nonlocal scale-scale correlation,
because the phase space cell of $(j+1, l+1)$ does not spatially
overlap with the cell $(j, [l/2])$. Therefore, the nonlocal
scale-scale correlation is measured by $C^{2,2}_{\bf j,
j+1}(\Delta l)$ with $|\Delta{\bf l}| \geq 1$.

Figure 9 plots $C^{2,2}_{\bf j, j+1}(\Delta l)$ vs. $\theta$ for
${\bf j}=(j,j)$ and $j=6$, 7, 8 and 9. The error bars are 90\%
confidence. For all panels, the data points from left to right
correspond, one by one, to $|\Delta{\bf l}| =0$, 1, 2... The
figure shows that the correlation $C^{2,2}_{\bf j, j+1}(\Delta l)$
for the 2MASS samples are larger than the random data on the first
two data points, but after the third data point, the 2MASS samples
do not show significant difference from the random data. Therefore,
the local scale-scale correlations ($|\Delta{\bf l}| =0$) are strong.
Non-local scale-scale correlations only exist between cells which
are spatially nearest neighbors, i.e.
$|\Delta{\bf l}| =1$. The scale-scale correlations with
$|\Delta{\bf l}|\geq 1$ generally are negligible. For $j=$ 7, 8 and
9, the spatial distance between nearest neighbors are only, respectively,
$28.3/2^j=0.22$, 0.11 and 0.06 angular degrees. On the other hand,
the SFCs have strong scale-scale correlation on, at least,
$|\Delta{\bf l}| =3$ (Fig. 6). In this sense, the scale-scale
correlation is also approximately spatially localized.

\section{Discussions and Conclusions}

The 2-D distribution of 2MASS galaxies has been studied using the
DWT, in which the SFC $\epsilon_{\bf j,l}$ is the count-in-cell in
physical space, while the WFC
$\tilde{\epsilon}_{\bf j,l}$ is a count-in-cell in phase space. We
find that the statistical properties of the SFC and WFC variables
are very different. The former is non-Gaussian and nonlocally
correlated, while the later is non-Gaussian, but its correlation
basically is spatially localized. That is, the sample has the following
statistical behavior in the phase space
\begin{itemize}
\item the one-point distribution of $\tilde{\epsilon}_{\bf j,l}$ is highly
   non-Gaussian on angular scales less than 1$^{\circ}$, corresponding to
   $\simeq 3.5$ h$^{-1}$ Mpc at the median redshift of the sample.
\item The 2nd and 4th order mode-mode correlations with
  $\tilde{\epsilon}_{\bf j,l}$ are always spatially localized. For highest
   $j$,
  the locality holds on angular scales as small as 0.06$^{\circ}$, or
  $\simeq$ 0.2 h$^{-1}$ Mpc at the median redshift of the sample.
\item the local scale-scale correlations are significant on scales less
      than 0.4$^{\circ}$, corresponding to $\simeq 1.5$ h$^{-1}$ Mpc at the
      median redshift of the sample. The scale-scale correlation is also
      approximately spatially localized.
\end{itemize}
A direct physical meaning of these results is that the cosmic
gravitational instability causes only strong interaction between the
modes in different wavebands and in the same spatial area, but
is weak for modes in different spatial area.

Because the nonlinear evolution of cosmic gravitational clustering has
the memory of its initial spatial correlation in the phase space,
the observed spatial locality of the 2MASS galaxies provides solid
evidence for models assuming that the initial perturbations are
spatially uncorrelated among phase space modes. This result is
consistent with the assumption that the initial perturbations are Gaussian.
Although the Gaussianity of the initial perturbations on large scales
has been extensively tested with the temperature fluctuations of the
cosmic microwave background radiation, the test given by the spatial
locality is effective to comoving scales as small as
$\simeq$ 0.2 h$^{-1}$ Mpc.

The spatial locality has been studied with samples of Ly-alpha forests
(Pando, Feng \& Fang, 2001). Since Ly-alpha forests refer to weakly
non-linear clustering, their spatial locality can be used to rule out
the models of initially non-Gaussian fields which are non-spatially
localized, but cannot do so for models of initial non-Gaussian which
are spatially localized. In this paper, the spatial locality is found
for sample referring to fully nonlinear regime. The memory of the
spatial locality is based on the halo model, for which an initially
Gaussian field is necessary. In this sense, the spatial locality of
the 2MASS gives stronger support to the initially Gaussian assumption
than that that of Ly-alpha forests.

The one-point distribution of 2MASS galaxies (Fig. 3) is very different
from that given by N-body simulation sample. The later in nonlinear regime
generally is lognormal (see Fig. 6 of Feng \& Fang, 2004), while the former
is more complicated (Fig. 3). This difference indicates that galaxy
distribution is biased with respect to the underlying dark matter field.
Nevertheless, the galaxy distribution is also highly spatially localized.
This result indicates that relationship between $\rho_g({\bf x})$ and
$\rho({\bf x})$ should be localized in phase space. That is, the
stochasticity and nonlocality of the relation between $\rho_g({\bf x})$
and $\rho({\bf x})$ are limited in the cell of phase space. For a
structure consisting of perturbations on
scales ${\bf k}$ to ${\bf k}+\Delta {\bf k}$, the nonlocal size in the
relation between $\rho_g({\bf x})$ and $\rho({\bf x})$ should not be larger
than $|\Delta {\bf x}|=2\pi/|\Delta {\bf k}|$. Otherwise, the galaxy bias
will violate the spatial locality of the galaxy distribution even when
the underlying field is spatial localized. Hence, from the 2MASS galaxies
we can conclude that the stochasticity and nonlocality of the bias
relation between $\rho_g({\bf x})$ and $\rho({\bf x})$ probably are no
more than that given by the uncertainty relation
$|\Delta {\bf x}||\Delta {\bf k}|=2\pi$.

\acknowledgments

The authors thank Dr. J. Pando for his assistance. This research is
partially supported by the National Basic Research Priorities Programmer
of The Ministry of Science and Technology of China.

\clearpage

\begin{figure}
\figurenum{1}\epsscale{1.0}\plotone{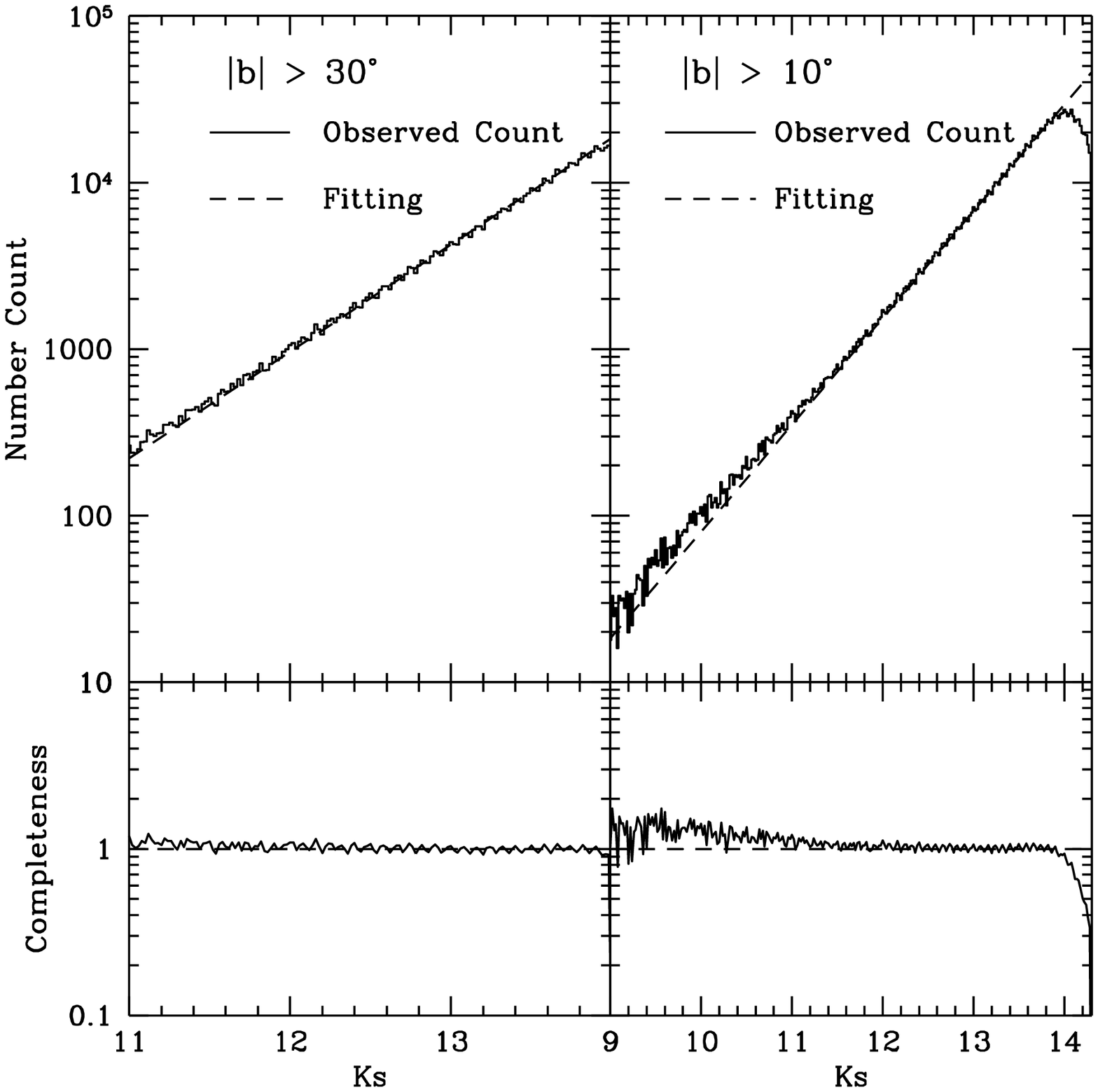} \figcaption{The number
counts $dN/dm$ and completeness $C(m)$ for 1.) the (standard)
sample of $|b| > 30^{\circ}$ and $12<K_s<13.7$ (left panels), and
2.) sample of $|b| > 10^{\circ}$ region when $9<K_s<14$ (right
panels). }
\end{figure}

\begin{figure}
\figurenum{2}\epsscale{1.0}\plotone{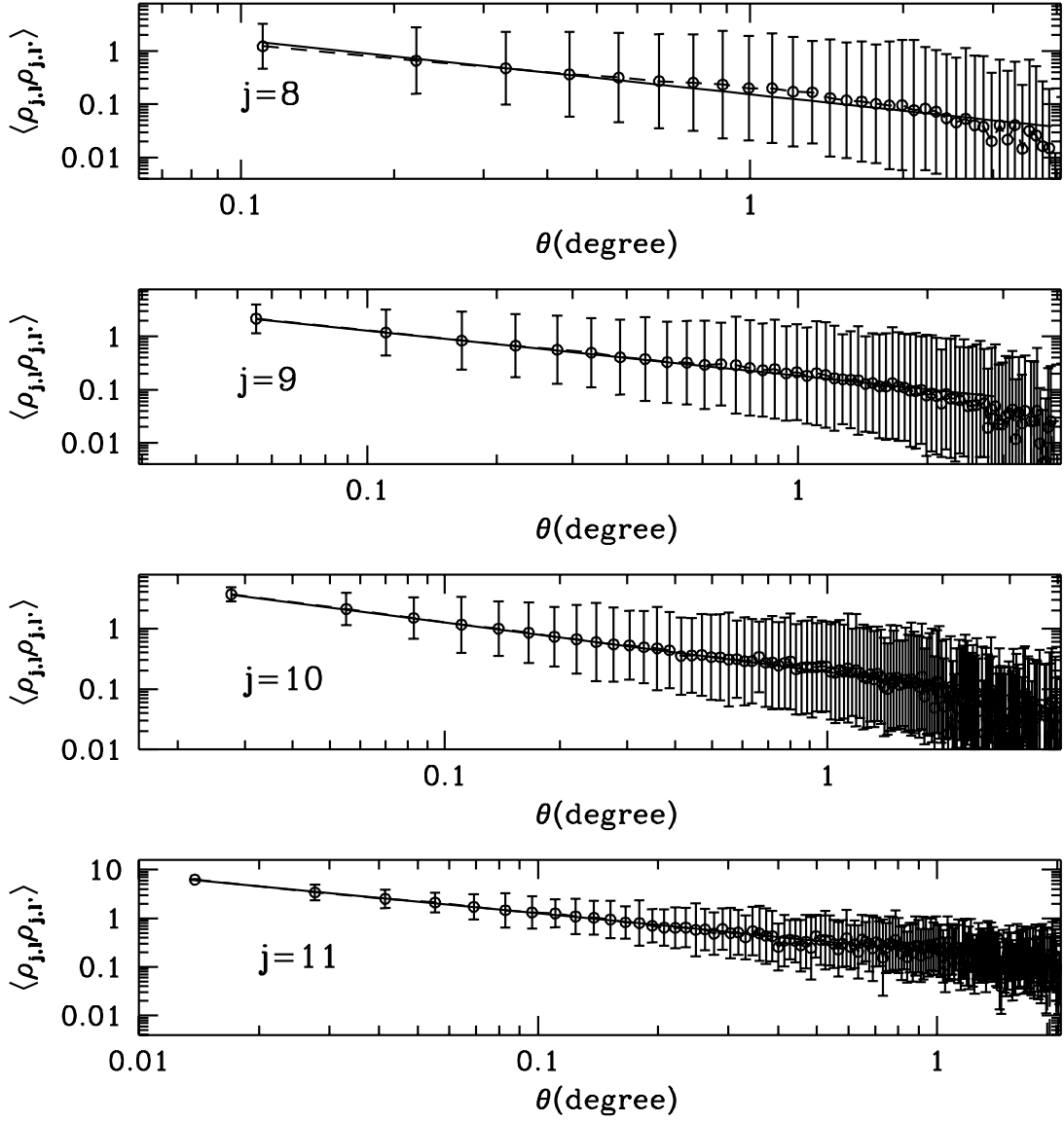} \figcaption{The SFC
correlation functions $\langle \rho_{\bf j,l}\rho_{\bf
j,l'}\rangle$ vs. $\theta$. The angular distance $\theta= |{\bf
l-l'}|\sqrt{800}/2^j$ angular degree. $j=8$, 9, 10 and 11, which
is the scales $\sqrt{800}/2$ of smoothing the data by the scaling
function. }
\end{figure}

\begin{figure}
\figurenum{3}\epsscale{1.0}\plotone{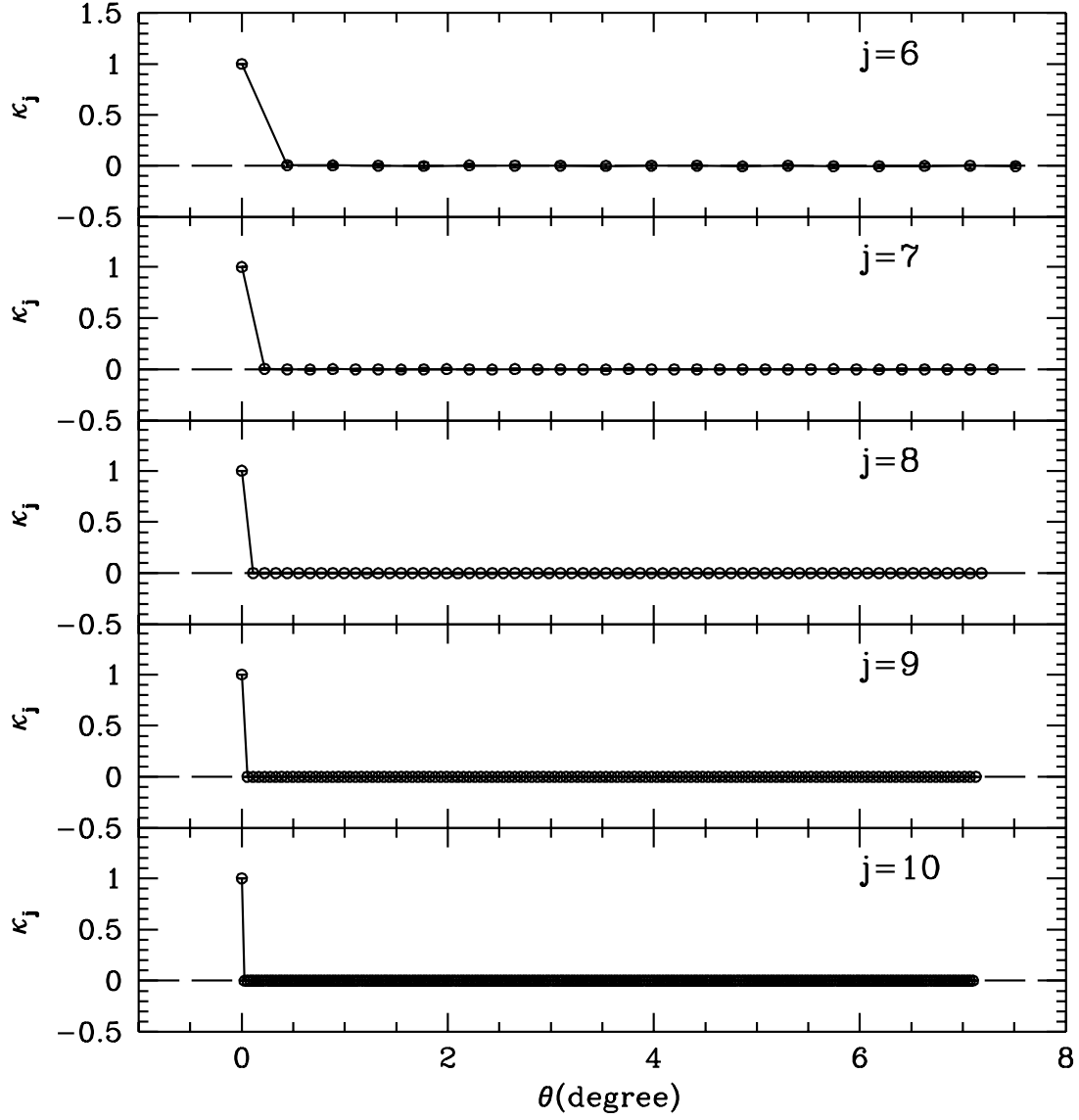} \figcaption{The WFC
correlation function $\kappa_{\bf j}$ vs. $\theta$. The angular
distance $\theta= |{\bf l-l'}|\sqrt{800}/2^j$ angular degree.
${\bf j}=(j,j)$ and $j =$ 5, 6, 7, 8, 9 and 10. }
\end{figure}

\begin{figure}
\figurenum{4}\epsscale{1.0}\plotone{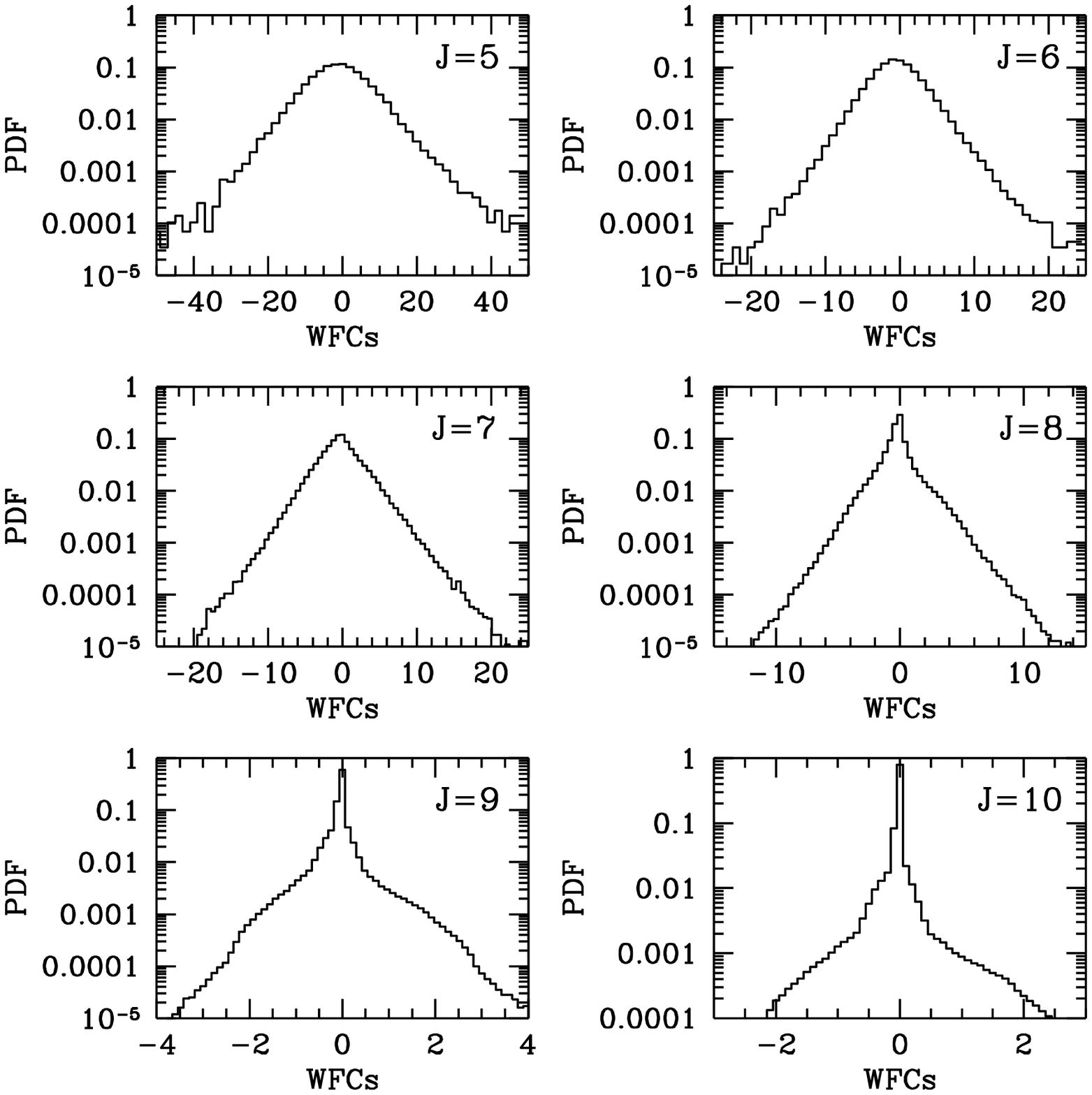} \figcaption{The
one-point distribution of WFCs $\tilde{\epsilon}_{\bf j, l}$ for
${\bf j}=(j,j)$ and $j =$ 5, 6, 7, 8, 9 and 10. }
\end{figure}

\begin{figure}
\figurenum{5}\epsscale{1.0}\plotone{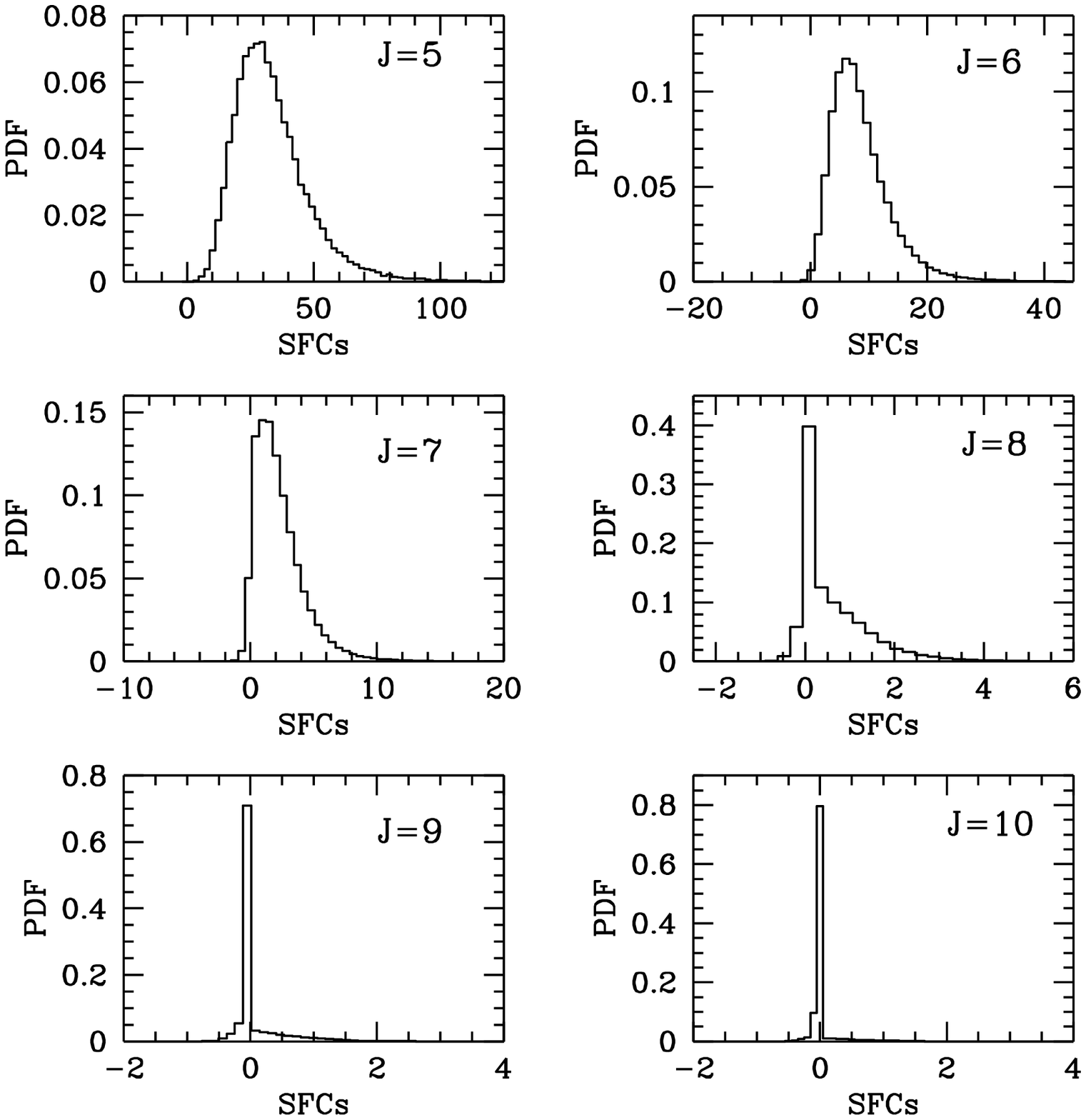}\figcaption{The
one-point distribution of SFCs $\epsilon_{\bf j, l}$ for ${\bf
j}=(j,j)$ and $j =$ 5, 6, 7, 8, 9 and 10. }
\end{figure}

\begin{figure}
\figurenum{6}\epsscale{1.0}\plotone{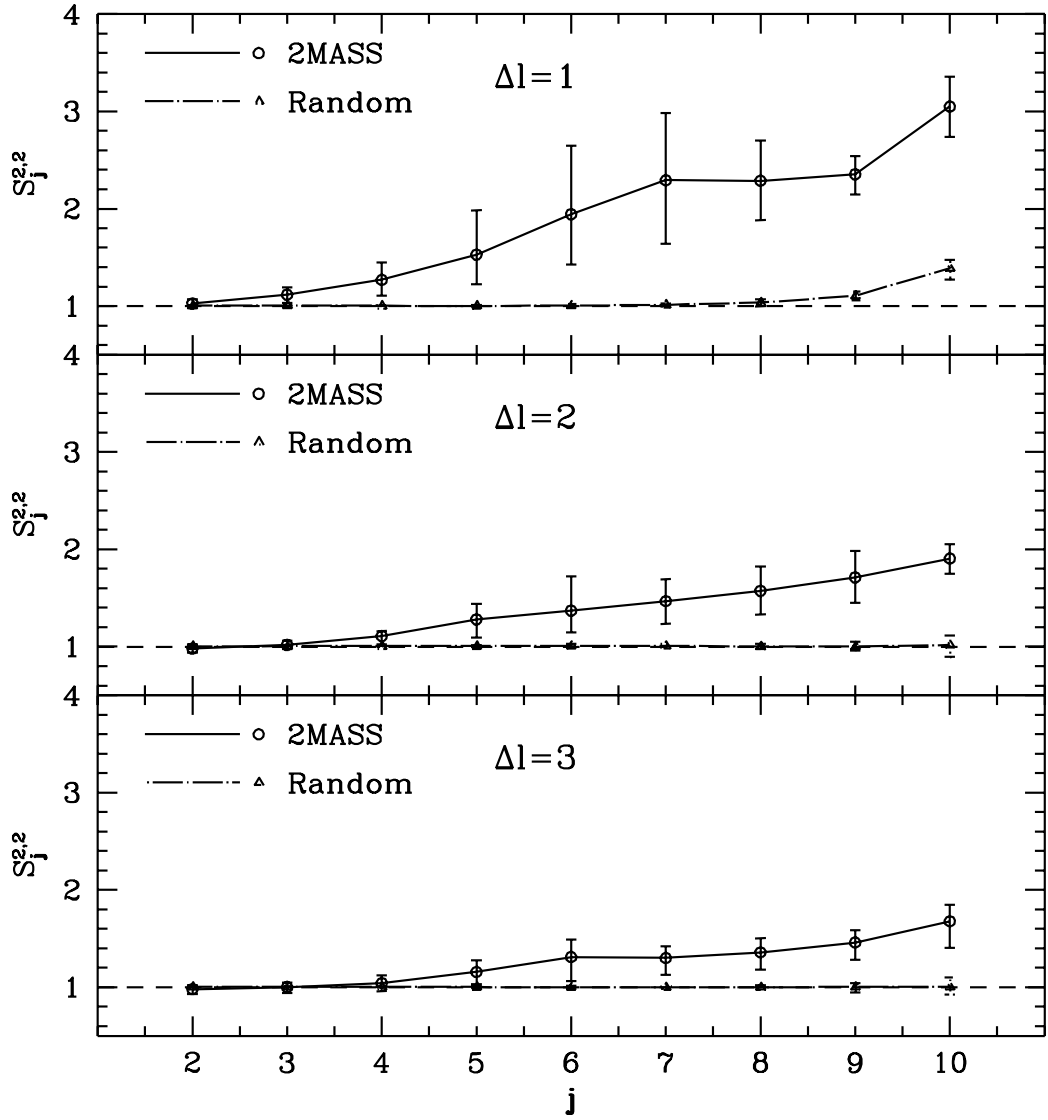} \figcaption{The
4$^{th}$ order statistics $S^{2,2}_{\bf j}$ vs. $j$ with $|\Delta
l|=1$, 2 and 3. The angular distance $\theta= |{\bf
l-l'}|\sqrt{800}/2^j$ angular degree. }
\end{figure}

\begin{figure}
\figurenum{7}\epsscale{1.0}\plotone{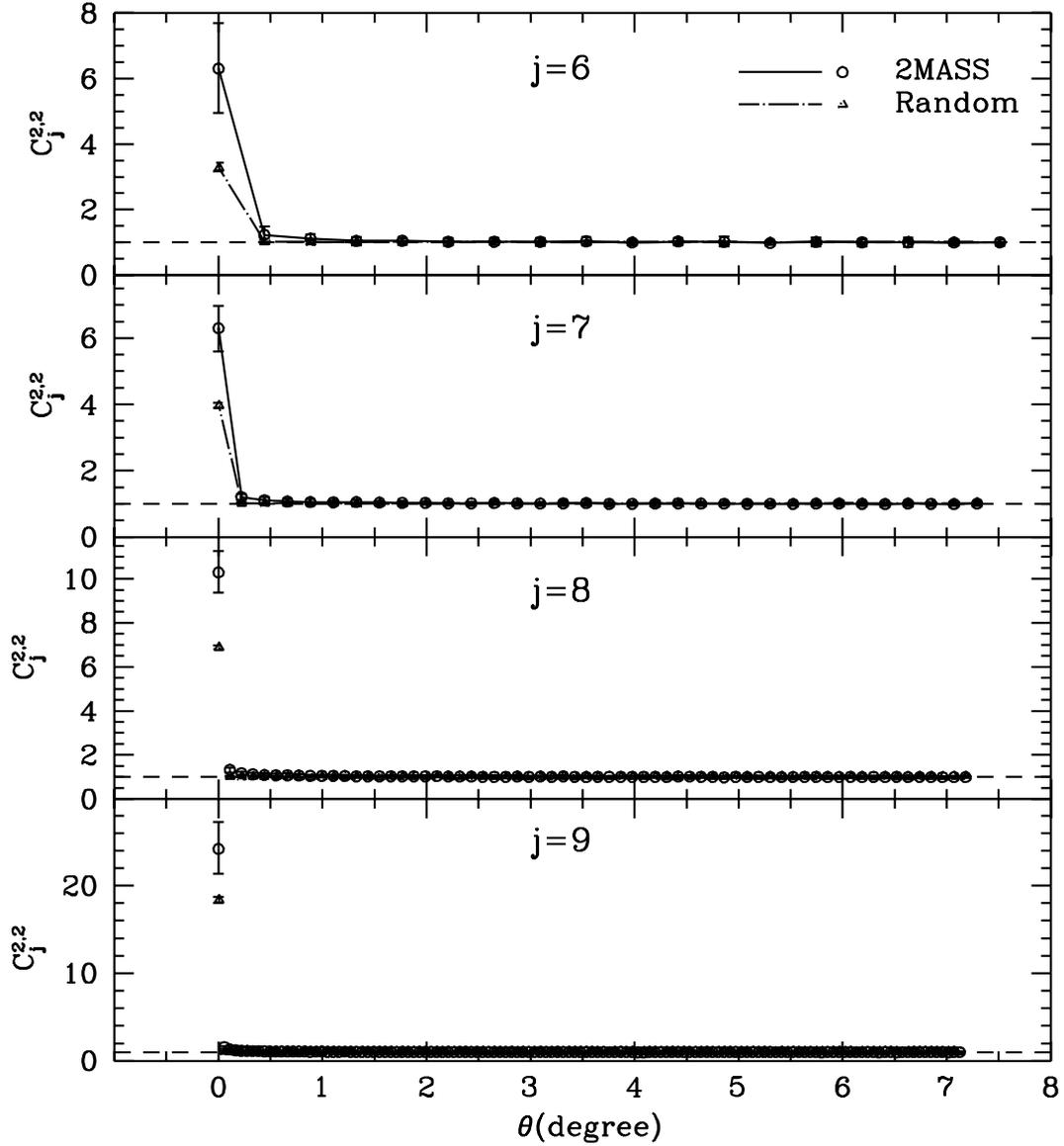} \figcaption{The
4$^{th}$ order statistics $C^{2,2}_{\bf j}$ vs. $\theta$. The
angular distance $\theta= |{\bf l-l'}|\sqrt{800}/2^j$ angular
degree. ${\bf j}=(j,j)$ and $j =$ 5, 6, 7, 8 and 9.}
\end{figure}

\begin{figure}
\figurenum{8}\epsscale{1.0}\plotone{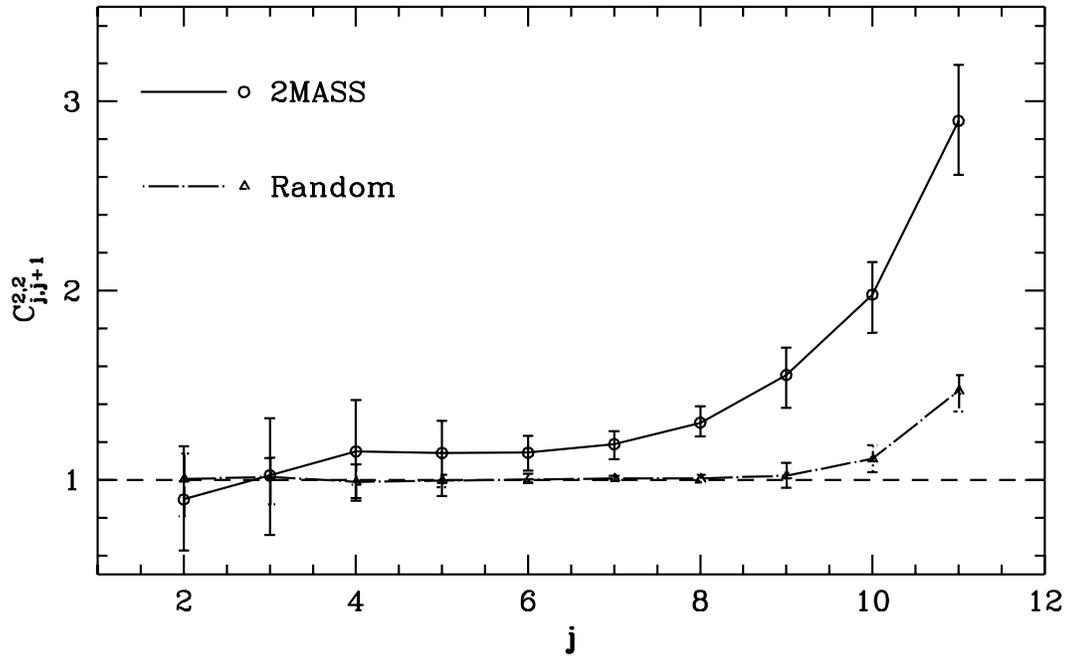} \figcaption{The local
scale-scale correlations $C^{2,2}_{\bf j,j+1}$ vs. $j$. The
angular scale of $j$ is $\sqrt{800}/2^j$ }
\end{figure}

\begin{figure}
\figurenum{9}\epsscale{1.0}\plotone{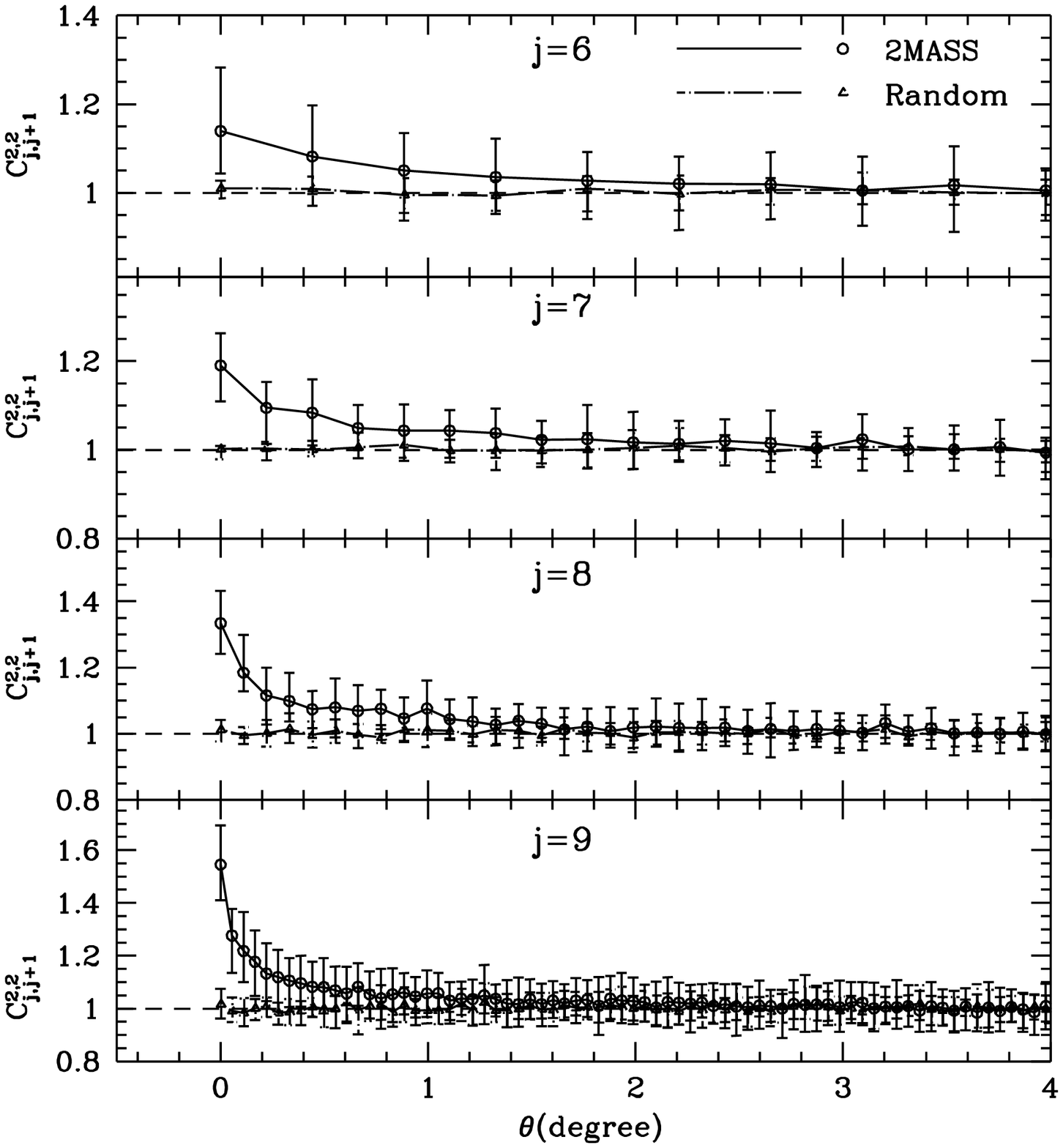}\figcaption{The
nonlocal scale-scale correlations $C^{2,2}_{\bf j,j+1}(\Delta l)$
vs. $\theta$. The angular distance $\theta= |\Delta {\bf
l}|\sqrt{800}/2^{j}$,  ${\bf j}=(j,j)$ and $j =$ 6, 7, 8, 9. }
\end{figure}


\begin{references}

\reference{} Afshordi, N., Loh, Y., \& Strauss, M.,
astro-ph/0308206

\reference{} Cole S. and Kaiser, N. 1988, \mnras, 233, 637

\reference{} Cooray, A. and Sheth, R. 2002, Physics Reports, 372, 1latex

\reference{} Daubechies I. 1992, Ten Lectures on Wavelets,
 (Philadelphia: SIAM)

\reference{} Dekel, A. \& Lahav, O. 1999, \apj, 520, 24

\reference{} Fang, L.Z. \& Feng, L.L. 2000, \apj,  539, 5

\reference{} Fang, L.Z. and Thews, R. 1998, Wavelets in Physics,
  World Scientific, (Singapore)

\reference{} Feng, L.L. \& Fang, L.Z. 2004, \apj, in press

\reference{} Frisch, U. 1995, Turbulence, (Cambridge Univ. Press)

\reference{} Greiner, M., Giesemann, J., Lipa, P., and Carruthers, P. 1996,
  Z. Phys. C69, 305

\reference{} Greiner, M., Lip, P., \& Carruthers, P., 1995,
 Phys. Rev. E51, 1948.

\reference{} Jarrett, T. et al. 2000, \aj, 119, 2498.

\reference{} Komatsu, E. et al. 2003, \apjs 148, 119

\reference{} Maller, A.H., McIntosh, D.H., Katz, N. \& Weinberg, M.D.
   astro-ph/03024005

\reference{} Matsubara, T. 1999, \apj, 525, 543

\reference{} Meneveau, C. \& Sreenivasan, K.R. 1987, Phys. Rev. Lett.
   59, 1424

\reference{} Pando, J. Feng, L.L.  \& Fang, L.Z. 2001, \apj, 554, 841

\reference{} Pando, J., Lipa, P., Greiner, M. and Fang, L.Z. 1998,
     \apj, 496, 9

\reference{} Pando, J., Valls-Gabaud, D. \& Fang, L.Z. 1998, Phys. Rev. Lett.,
   81, 4568

\reference{} Peebles, P. 1980, The large scale structures of the universe,
    (Princeton press)

\reference{} Press, W.H., Teukolsky, Vetterling, W.T. \& Flannery,
    B.P. 1992, Numerical Recipes, (Cambridge Univ. Press)

\reference{} Soneira, R.M. \& Peebles, P.J.E. 1977, \apj, 211, 1

\reference{} Strauss, M.A., Ostriker, J.P. \& Cen, R. 1998, \apj,
  494, 20

\reference{} Yang, X. H., Feng, L.L., Chu, Y.Q., \& Fang, L.Z.
2001, \apj, 560, 549

\end{references}
\end{document}